\documentclass{icrc}

\usepackage{times}
\usepackage{graphicx} 

\begin{document}

\title{Measurement of the Cherenkov light spectrum and of the polarization with the HEGRA-IACT-system}
\author[1]{M. D\"oring}
\author[1]{K. Bernl\"ohr}
\author[1]{G. Hermann}
\author[1]{W. Hofmann}
\author[1]{H. Lampeitl}
\affil[1]{Max-Planck-Institut f\"ur Kernphysik Heidelberg, Postfach 103980, 69029 Heidelberg, Germany}

\correspondence{\\M. D\"oring  \\     
(Michael.Doering@mpi-hd.mpg.de)}

\firstpage{1}
\pubyear{2001}


\maketitle

\begin{abstract}
The HEGRA system of Imaging Atmospheric \linebreak Cherenkov Telescopes (IACTs) 
detects Cherenkov light produced by air showers. 
The concept of stereoscopic observation with the five HEGRA telescopes
allows the reconstruction of various shower parameters, 
for example the shower direction, the location of the shower core and 
the energy of cosmic rays.
One of the telescopes was modified so that measurements of the spectrum 
and the polarization of Cherenkov light with the HEGRA system were
possible. The experimental setup is described and preliminary results presented.
\end{abstract}

\section{Introduction}
Imaging atmospheric Cherenkov telescopes (IACTs) have \linebreak emerged as the prime
instruments to detect cosmic gamma rays in the TeV energy range. A key
area of current research with IACTs are the precision measurements of
energy spectra of AGNs, the study of cutoffs in the spectra (see, e.g. 
Kohnle 2001), and their interpretation in terms of
interactions with infrared/optical background fields. For these
investigations an absolute energy calibration is important, but lacking a
``calibration beam'', this can only be done in indirect ways, highly
depending on Monte Carlo simulations.  The detected \linebreak Cherenkov light yield
from cosmic-ray induced air showers combines Cherenkov emission by the
shower particles, the atmospheric extinction, the mirror reflectivity and
photon detector efficiencies. With the significant uncertainties involved
in the modeling, there is a clear need to test all aspects of the
simulations to ensure reliability.  Some properties, such as the
distribution of IACT image shapes, have been studied extensively.  The
radial distribution of Cherenkov light in gamma-induced showers was
measured by HEGRA \cite{ah99a}. In this work, we discuss
measurements of two other key characteristics of Cherenkov light: its
spectrum and its polarization.
The measurements were performed with the \linebreak HEGRA system of IACTs
\cite{da97} .  Four of the five telescopes were used to provide a trigger,
to select air showers and to define the shower characteristics. The fifth
telescope was equipped either with different optical filters or with a
polarization filter and served to analyze the characteristics of the
Cherenkov light. Since the fifth telescope is not participating in the
selection or definition of events, the effect of adding filters or
polarizers can be studied in a unbiased way, by comparing the average
intensity detected in the image with and without the optical elements.

\section{Cherenkov spectrum and polarization}
The number of emitted Cherenkov photons per wavelength intervall  is proportional to
$1/\lambda^2$. Wavelength dependent attenuation in the atmosphere leads
to the spectrum pictured in Fig.1. 
\begin{figure}[h]
\vspace*{2.0mm} 
\includegraphics[width=8.3cm]{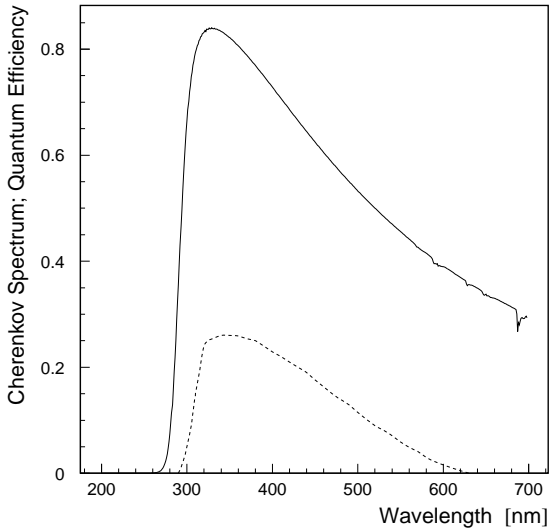}
\caption{Typical spectrum of Cherenkov light generated by vertical
TeV air showers at the altitude of the HEGRA IACT system, 2200 m
asl (in arbitrary units).
Shown as a dashed line is the quantum efficiency of the type of 
photomultipliers used in the HEGRA cameras.}
\end{figure}
Depending on the exact atmospheric
model and the assumptions concerning the distribution of aerosols,
in particular the short-wavelength end of the spectrum can vary
considerably \cite{be99}. To obtain the detected spectrum,
this spectrum has to be folded with the PMT quantum efficiency, which again
shows significant uncertainties and variations between production batches.
Cherenkov radiation is naturally 100\% linearly polarized. The
polarization vector is perpendicular to the Cherenkov cone, pointing away
from the particle's path. Multiple scattering in the shower causes an
angular distribution of secondaries relative to the shower axis, resulting
in a dilution of the net polarization observed on the ground. One should
expect that the degree of polarization exhibits a maximum near the
Cherenkov radius of about 120 m, where radiation from the well-collimated
upper part of the shower is collected, and decreases both for smaller core
distances (for symmetry reasons, light observed on the shower axis has to
be unpolarized) and for larger core distances, where heavily scattered
particles contribute.  Ideas concerning the
use of the polarization information in the
IACT technique as well as simulations of the polarization properties of
Cherenkov light from air showers were presented
by Hillas (1996),
by Contreras et al. (98), and also by Gokhale et al. (2001).
A significant degree of
polarization was pointed out, with a peak at $\approx 25 - 32\%$ for
primary gamma rays and core distances from 60~m to 140~m, and $\approx 22
- 26 \%$ for protons in a similar distance range. 
 Experimental results for cosmic rays with primary energies $\ge 4$~PeV were given by
\cite{ti99}. An average degree of polarization of
40\% was reported, with little variation with the distance to the shower core.

\section{Measurements}
The measurements were realized with the HEGRA Cherenkov telescopes.  The
HEGRA system of Imaging Cherenkov Telescopes (IACTs) is located on the
Canarian Island La Palma (at $28^{\circ}$45' N, $17^{\circ}$ 53 W) at 2200m
a.s.l.  It consists of five Cherenkov Telescopes (CT2 - CT6) , four in the
corners of a square with ~100~m side length and one central telescope
(CT3). The stereoscopic detection of Cherenkov light produced by
cosmic-ray induced extensive air showers with an IACT system allows the
reconstruction of shower parameters  such as the
location of the shower core, the shower direction and the energy of the
primary particle \cite{da97}. Each of the HEGRA telescopes has a
field of view of $4.3^{\circ}$. The cameras consist of 271
photomultipliers, each with a field of view of $0.25^{\circ}$
The measurements of the Cherenkov light spectrum were realized by using
optical glas filters.  Two different sets of filters were used:\newline 
(a) bandpass filters with center wavelengths at 325~nm (80~nm FWHM), 
380~nm (120~nm~FWHM), 440~nm (100~nm FWHM) and 515~nm (65~nm FWHM),
see Fig.2.
\vspace{0.3cm}
\newline
(b) long pass edge filters with cut-off wavelengths at 350~nm, 370~nm, 408~nm,
450~nm and with an average transmission of 80-90\% (see Fig.3.).\\

\begin{figure}[t]
\vspace*{2.0mm} 
\includegraphics[width=8.3cm]{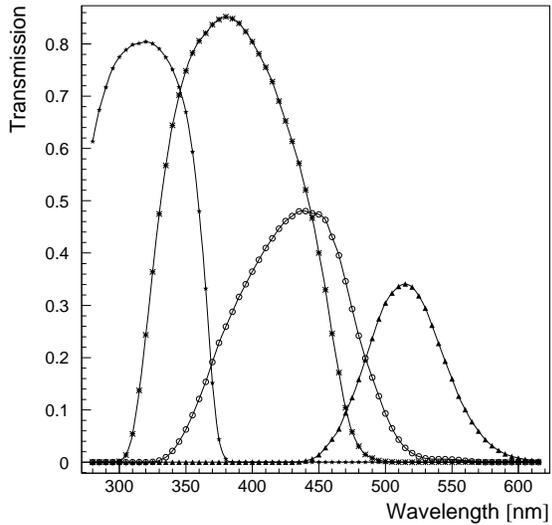}
\caption{Transmission curves of the four bandpass filters as measured
  in the laboratory}
\end{figure}
\begin{figure}[t]
\includegraphics[width=8.3cm]{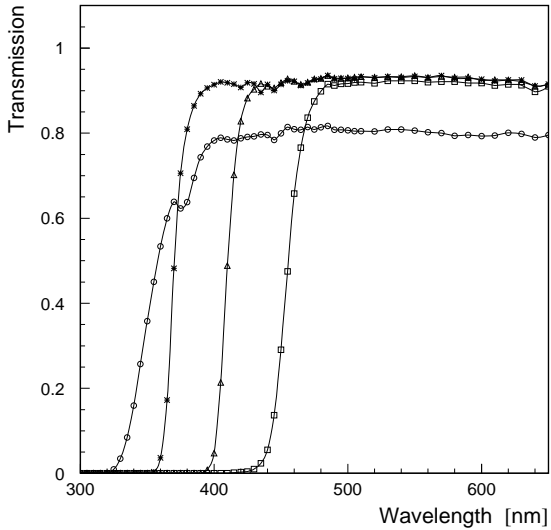}
\caption{Transmission curves of the four long pass edge filters as
  measured in the laboratory}
\end{figure}
Since filters were only available as
squares with 50~mm side length, many filter elements were mounted on an
aluminum grid such as to cover the entire active area of the camera of the
central telescope CT3. The grid partially obscures some of the camera
pixels (see Fig.4.).
\begin{figure}[h]
\vspace*{2.0mm} 
\includegraphics[width=8.3cm]{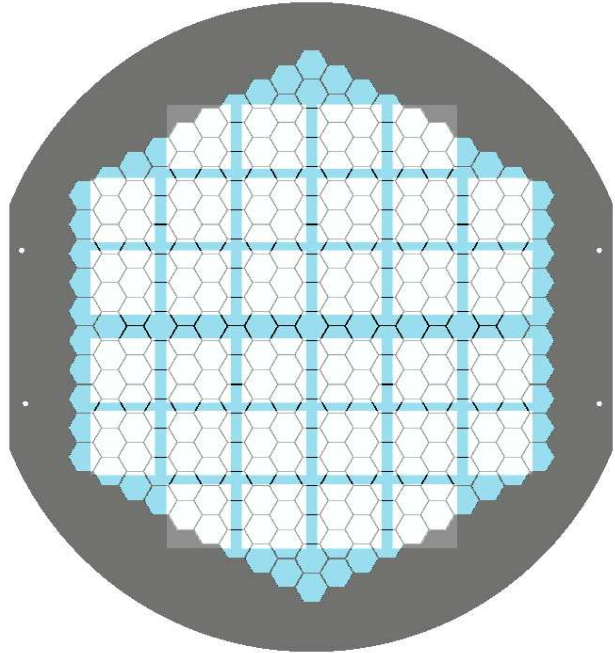} 
\caption{The aluminium grid is drawn in grey, the filters are white
  squares. The hexagonal structure is the pixel matrix.}
\end{figure}
The different filters of type (a) or (b) were measured
simultaneously by dividing the camera into four quadrants, and using a
different filter wavelength in each quadrant. For the study of the
spectrum, 14 hours of data were taken, with the telescopes pointing to a dark
area of the sky near the zenith ($\theta_z \approx 6^\circ ... 8^\circ$).
To test systematics, the filter plate was rotated between sets of
measurements, and reference measurements were taken both with the empty
filter holder, and with the unobscured camera.
For the polarization measurements the linear polarizing filter Schott LP32
was used. Its transmission for light polarized in the filter direction
varies between 20\% and 75\% depending on wavelength as measured in the 
laboratory; transmission for polarization perpendicular to the filter
direction is less than 2\% of the transmission in filter direction, 
for all wavelengths. For the polarization studies 5 h of data were taken,
again pointing with the telescopes to the zenith.

\section{Data Analysis: Spectrum}
Since data were taken for cosmic rays near the zenith, without
specifically aiming at a gamma-ray source, all results refer to
nucleon showers.
For the analysis, events were selected where all 4 of the outer
telescope triggered and provide useful images. Based on these images, the
shower direction and energy are reconstructed, and a prediction for the
center of gravity and image orientation in the central telescope CT3 was
derived. In CT3, pixels were then selected within a distance of $0.5^\circ$
from the predicted images axis. This cut keeps virtually all pixels
illuminated by the air shower. Only those pixels were used in the
analysis, which were not obscured by the filter support structure, and
which were illuminated through filters of the same quadrant (i.e., of the
same wavelength). Hence, only 92 out of 271 pixels were used in the spectrum
analysis. Pixel signals were normalized to the reconstructed shower
energy.  For the current analysis, the light detected in all pixels with
filters of a given wavelength was summed up, and the light yield was
normalized to the yield without filters. In a future publication, the
dependence on the location of a pixel within the image (in the head or
tail end of the shower, on or off the shower axis) will be addressed.
In Table 1. and Table 2. (preliminary) mean values of the amplitude for each
filter are shown, normalized to the yield without filters. The quoted
errors are statistical errors. They are compared with calculations,
which include the atmospheric extinction, the mirror reflectivity,
filter transmission and the wavelength dependent quantum efficiency of
the photomultipliers. Here the errors were estimated through
combining different aerosol models, namely navy maritime,
maritime, and rural as well as slightly different tables for the
quantum efficiency of the photomultipliers.
\begin{table}[h]
\begin{tabular}[h]{|c|c|c|} \hline
center &  measured & calculated \\
 wavelength [nm]  & intensity [\%]  & intensity [\%]  \\ \hline
325nm & 28.5 $\pm$ 0.5 & 28.2 $\pm$ 2.9\\ \hline
380nm & 51.5 $\pm$ 0.9 & 50.5 $\pm$ 3.0\\ \hline
440nm & 22.0 $\pm$ 0.4 & 19.0 $\pm$ 1.0\\ \hline
515nm &  4.6 $\pm$ 0.1 &  3.3 $\pm$ 0.2\\ \hline
\end{tabular}
\caption{Relative intensities measured with the bandpass filters}
\end{table}
\begin{table}[h]
\begin{tabular}[h]{|c|c|c|}\hline
cut-on & measured & calculated \\
wavelength [nm] & intensity [\%] & intensity [\%] \\ \hline
350nm & 53.4 $\pm$ 1.5 & 54.0 $\pm$ 3.0\\ \hline
370nm & 51.7 $\pm$ 1.5 & 53.7 $\pm$ 3.1\\ \hline
408nm & 36.6 $\pm$ 1.0 & 34.2 $\pm$ 2.3\\ \hline
450nm & 20.1 $\pm$ 0.6 & 18.2 $\pm$ 1.5\\ \hline
\end{tabular}
\caption{Relative intensities measured with long pass edge filters}
\end{table}

\section{Data Analysis: Polarization}
The basic analysis procedure resembles the one for the spectral analysis;
to determine the degree of polarization, the \linebreak summed light
yield of all image pixels is normalized to the shower
energy, and is plotted as a
function of the angle $\theta$ between the image of the shower axis and
the transmission direction of the polarizer
(note that the image of the
shower axis points to the shower core). To study the dependence of the
polarization on the distance $r$ between central telescope and shower core
position, the distance was divided into intervals of 20~m, from 10~m to
170~m. Fig.5. shows the mean of the energy normalized amplitude as a
function of $\theta$ for 50~m~$< r <$~70~m. For angles of $0^\circ$ and
$180^\circ$ polarization axis of the filter and main axis of the ellipse
are parallel. The degree of polarization is calculated as
\begin{equation}
p =
\frac{I_{max}-I_{min}}{I_{max}+I_{min}}\ \label{eq_poldegr}
\end{equation}
where $I_{max}$ and $I_{min}$ are the maximum and minimum amplitudes,
determined by fitting the $\theta$-depedence with \linebreak
$a + b \cos(2\theta)$.
The resulting polarization p(r) = b/a is shown in Fig.6.
It shows the expected linear rise for core distances below 100~m,
and decreases for core distances larger than 130~m.
\begin{figure}[t]
\includegraphics[width=8.3cm]{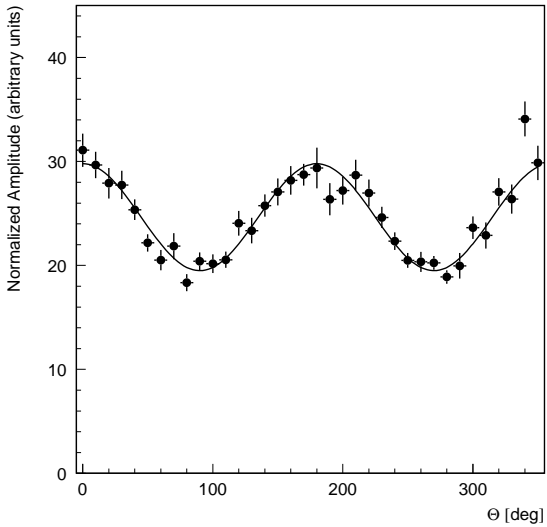} 
\caption{Energy normalized amplitude as a function of $\theta$ for a
  distance to the core postion of r~=~50~m...70~m. The solid line
  indicates a fit to determine the degree of polarization.}
\end{figure}
\begin{figure}[t]
\includegraphics[width=8.3cm]{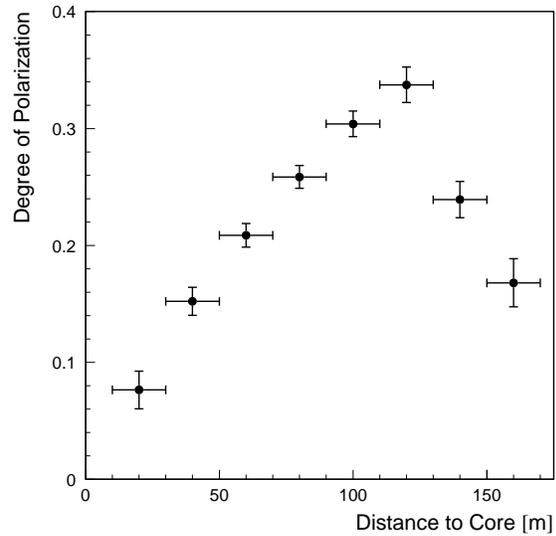} 
\caption{Degree of polarisation measured as a function of the core distance}
\end{figure}
\section{Conclusion}
The stereoscopic observation of hadron induced air showers allows the
measurement of the Cherenkov light spectrum and its polarization.
The first results of the spectrum measurements  are in good agreement 
with calculations, taking into account the atmospheric extinction, mirror 
reflectivity, quantum efficiencies of the photomultipliers and 
filter transmissions. 
First results of the polarization analysis show the expected
dependance on the distance to the shower core.
A more detailed analysis is in progress.\newline
\newline \newline 

\begin{acknowledgements}
The support of the HEGRA experiment by the German Ministry for
Research and Technology BMBF and by the Spanish Research Council CYCIT
is acknowledged. We are greatful to the Instituto de Astrofisica de
Canarias for the use of the site and for providing excellent working
conditions. We thank the other members of the HEGRA CT group, who
participated in the construction, installation, and operation of the
telescopes. We gratefully acknowledge the technical support staff of
Heidelberg, Kiel, Munich, and Yerevan.
\end{acknowledgements}

\end{document}